\documentclass[aps,prl,twocolumn,superscriptaddress,longbibliography]{revtex4-1}

\usepackage{graphicx}
\usepackage{amsmath}
\usepackage[normalem]{ulem}
\usepackage[usenames,dvipsnames]{xcolor}
\usepackage[colorlinks=true,linkcolor=blue,urlcolor=blue,citecolor=blue]{hyperref}
\usepackage{braket}
\usepackage{amssymb}
\usepackage{commath}
\usepackage{csquotes}
\usepackage{dcolumn}
\usepackage{bm}
\usepackage{epstopdf}
\usepackage{lipsum}
\usepackage{tikz}

\newcommand*\circled[1]{\tikz[baseline=(char.base)]{
		\node[shape=circle,draw,inner sep=0.5pt] (char) {#1};}}

\newcommand{\eqw}[1]{(\ref{#1})}
\newcommand{\eq}[1]{Eq.~(\ref{#1})}
\newcommand{\fig}[1]{Fig.\thinspace{}\ref{#1}}
\newcommand{\fc}[1]{{#1}}
\newcommand{\figc}[2]{Fig.\thinspace{}\ref{#1}\thinspace{}\fc{#2}}
\newcommand{\cohimb}{$\widetilde{\mathcal{I}}$}

\pdfoutput=1

\begin{document}

\title{Probing Slow Relaxation and Many-Body Localization in Two-Dimensional Quasi-Periodic Systems}

\author{Pranjal Bordia}
\affiliation{Fakult\"at f\"ur Physik, Ludwig-Maximilians-Universit\"at M\"unchen, Schellingstr. 4, 80799 Munich, Germany}
\affiliation{Max-Planck-Institut f\"ur Quantenoptik, Hans-Kopfermann-Str. 1, 85748 Garching, Germany}

\author{Henrik L\"uschen}
\affiliation{Fakult\"at f\"ur Physik, Ludwig-Maximilians-Universit\"at M\"unchen, Schellingstr. 4, 80799 Munich, Germany}
\affiliation{Max-Planck-Institut f\"ur Quantenoptik, Hans-Kopfermann-Str. 1, 85748 Garching, Germany}

\author{Sebastian Scherg}
\affiliation{Fakult\"at f\"ur Physik, Ludwig-Maximilians-Universit\"at M\"unchen, Schellingstr. 4, 80799 Munich, Germany}
\affiliation{Max-Planck-Institut f\"ur Quantenoptik, Hans-Kopfermann-Str. 1, 85748 Garching, Germany}

\author{Sarang Gopalakrishnan}
\affiliation{Department of Engineering Science and Physics, CUNY College of Staten Island, Staten Island, NY 10314 USA}

\author{Michael Knap}
\affiliation{Department of Physics, Walter Schottky Institute, and Institute for Advanced Study, Technical University of Munich, 85748 Garching, Germany}

\author{Ulrich Schneider}
\affiliation{Fakult\"at f\"ur Physik, Ludwig-Maximilians-Universit\"at M\"unchen, Schellingstr. 4, 80799 Munich, Germany}
\affiliation{Max-Planck-Institut f\"ur Quantenoptik, Hans-Kopfermann-Str. 1, 85748 Garching, Germany}
\affiliation{Cavendish Laboratory, Cambridge University, J.J. Thomson Avenue, Cambridge CB3 0HE, United Kingdom}

\author{Immanuel Bloch}
\affiliation{Fakult\"at f\"ur Physik, Ludwig-Maximilians-Universit\"at M\"unchen, Schellingstr. 4, 80799 Munich, Germany}
\affiliation{Max-Planck-Institut f\"ur Quantenoptik, Hans-Kopfermann-Str. 1, 85748 Garching, Germany}

\date{\today}

\begin{abstract}
In a many-body localized (MBL) quantum system, the ergodic hypothesis breaks down completely, giving rise to a fundamentally new many-body phase. Whether and under which conditions MBL can occur in higher dimensions remains an outstanding challenge both for experiments and theory. Here, we experimentally explore the relaxation dynamics of an interacting gas of fermionic potassium atoms loaded in a two-dimensional optical lattice with different quasi-periodic potentials along the two directions. We observe a dramatic slowing down of the relaxation for intermediate disorder strengths and attribute this partially to configurational rare-region effects. Beyond a critical disorder strength, we see negligible relaxation on experimentally accessible timescales, indicating a possible transition into a two-dimensional MBL phase. Our experiments reveal a distinct interplay of interactions, disorder, and dimensionality and provide insights into regimes where controlled theoretical approaches are scarce.
\end{abstract}

\maketitle
\newpage
\textbf{\textit{Introduction:}} The ergodic hypothesis underlies quantum statistical mechanics, linking reversible microscopic dynamics to irreversible macroscopic behavior. In an ergodic system local degrees of freedom get rapidly entangled with one another and local quantum correlations are rapidly erased~\cite{Deutsch91, Srednicki94, Rigol08,Popescu2009dynamic}. Non-ergodic many-body localized (MBL)~\cite{Basko06, Polyakov05, Imbrie14, Nandkishore15, Altman15, Schreiber15} systems, however, defy this ubiquitous behavior and show persistent local quantum correlations~\cite{huse_pheno_14,serbyn_local_2013,serbyn_interferometric_2014,bahri_localization_2015}. Furthermore, MBL systems are believed to be robust to small, local perturbations and form a distinct, non-ergodic phase of matter. The phase transition from the ergodic phase to the MBL phase appears to be a highly unusual critical phenomenon; as ergodicity breaks down in the MBL phase, its description lies beyond the scope of thermodynamics and traditional statistical physics~\cite{Nandkishore15,Altman15}. 

\begin{figure}[t!]
	\centering
	\includegraphics[width=0.96\linewidth]{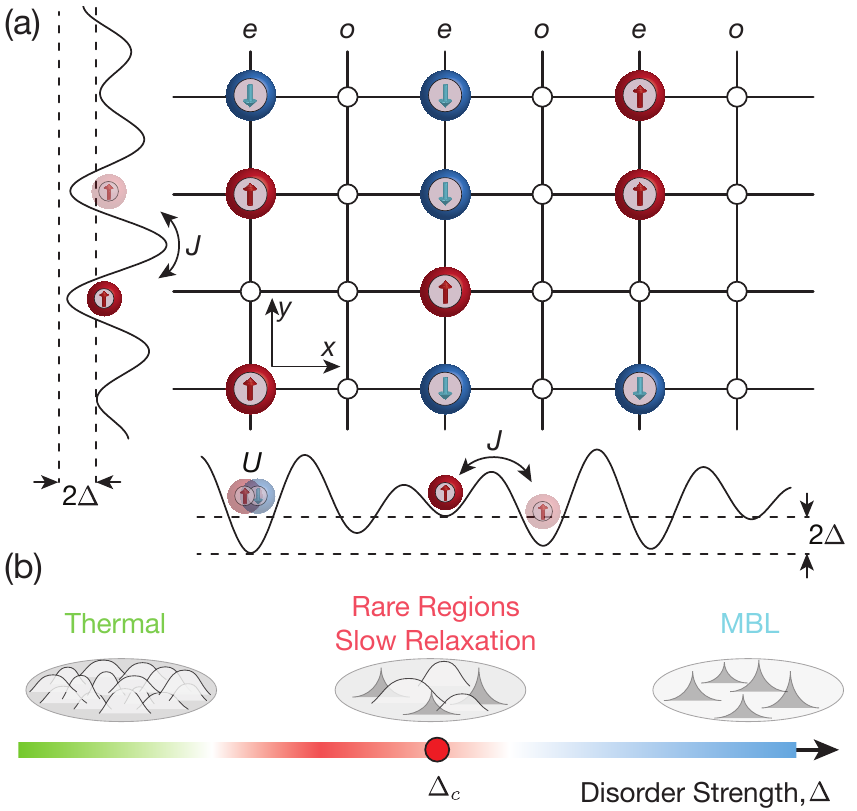}
	\caption{Schematic of the experiment. (a) The system is initialized in a striped density-wave pattern of fermionic ${}^{40}$K atoms in a random mixture of two spin states (red and blue) in a square lattice with tunneling matrix element $J$, quasi-periodic potential of strength $\Delta$, and tunable onsite interactions $U$ between the different spins. The largest realized 2D system is composed of approximately $200 \times 100$ sites with several thousand atoms. (b) For weak disorder strength the system thermalizes quickly (green area), whereas at strong disorder it is likely to exhibit a many-body localized regime (blue). Close to the transition (red dot, $\Delta_c$), a regime of slow relaxation is observed, potentially caused by locally insulating regions (red area).}
	\label{schematic}
\end{figure}

Due to limitations of the available numerical methods, most theoretical explorations of MBL concentrate on one-dimension. Whether, and under which conditions, MBL can occur in higher-dimensional systems remains a challenging question for both theory and experiment. While the initial theoretical work in Ref.~\cite{Basko06} on MBL does not depend strongly on dimensionality, it was recently argued that rare, locally thermal regions~\cite{deroeck2016} in systems with true random disorder, can destabilize the MBL phase in two dimensions. It is presently unclear if such arguments also hold for systems with deterministic disorder such as quasi-periodic potentials. At the same time, initial experiments provided evidence for an MBL phase in higher dimensions by measuring global transport~\cite{Kondov15, Choi16}. Moreover, the nature of a possible MBL transition in higher dimensions might itself be very different as compared to the one-dimensional transition~\cite{Agarwal15, Vosk_Theory_2015,  Potter_Universal_2015, Gopalkrishnan15, gopalakrishnan_griffiths_2016,agarwal_rr_16}; for example, a sub-diffusive phase as a precursor to localization in one-dimension~\cite{agarwal_rr_16, Luitz16Review, Agarwal15, znidaric_diffusive_2016, lueschencrit16,prelovrev16} might not exist in higher dimensions (but see Ref.~\cite{blr2}). Given the apparent conflict of available theoretical results~\cite{Chandran16, Inglis16, deroeck2016}, and infeasibility of reliable numerical simulations, experiments stand to play an important role in elucidating these regimes~\cite{Kondov15, Schreiber15, Smith2016, Choi16,Kucsko16}. 

Ultracold atoms in optical lattices provide a particularly well-suited platform to explore these phenomena, as they combine almost ideal isolation from the environment with individual experimental control of all microscopic parameters. In this work, we employ ultracold fermions in a quasi-periodic optical lattice to experimentally investigate the appearance of a non-ergodic many-body phase in two dimensions by directly tuning the strength of a quasi-periodic potential. By quantifying the dynamical relaxation of an imprinted striped density-wave pattern, we find evidence for three dynamical regimes: a regime of fast relaxation at weak disorders consistent with thermalization, a regime of slow relaxation at intermediate disorders, resembling the relaxation expected in a Griffiths regime~\cite{gopalakrishnan_griffiths_2016}, and, finally, a strong-disorder regime with negligible relaxation, consistent with the appearance of an MBL phase. The slow relaxation regime only begins once the single-particle states are already strongly localized, highlighting that the slow dynamics is an inherent interaction effect. The relaxation dynamics in two dimensions are distinctly different from the corresponding one-dimensional case~\cite{lueschencrit16,gopalakrishnan_griffiths_2016}, revealing the important role of dimensionality. Furthermore, tracking the relaxation dynamics appears to be useful in locating the many-body localization transition, even in the presence of weak couplings to the environment~\cite{Luitz16,lueschencrit16}.

\begin{figure*}[t!]
	\centering
	\includegraphics[width=1.0\linewidth]{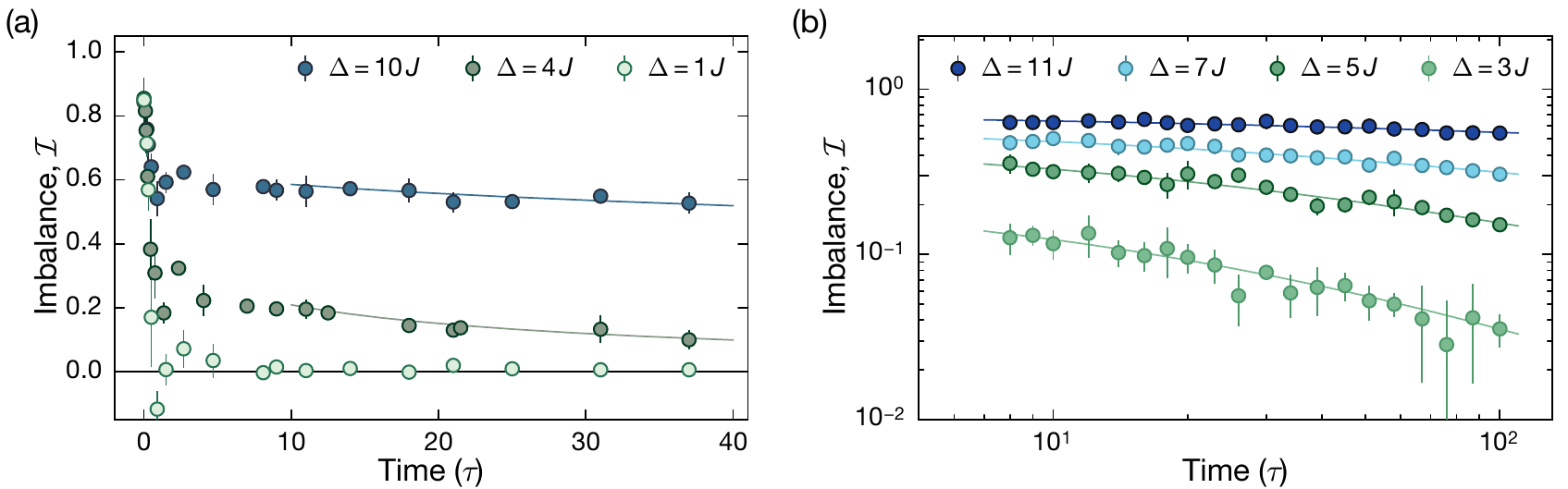}
	\caption{Time evolution of an imprinted density-wave pattern in the interacting, two-dimensional Aubry-Andr\'e model: We measure the time evolution of the imbalance between atom numbers on even and odd stripes for intermediate interactions $U=5\,J$ and varying disorder strength $\Delta$. (a) At weak disorder ($\Delta = 1\,J$) the imbalance vanishes quickly within a couple of tunneling times, signaling ergodic dynamics. For intermediate disorder ($\Delta = 4\,J$), we observe a markedly slow relaxation. At even stronger disorders ($\Delta =10\,J$), relaxation is absent up to a weak, previously measured~\cite{Bordia16} coupling to the environment. (b) The same time evolution is shown on a double logarithmic scale for additional values of the disorder strength. The solid lines denote fits to the model described in the main text ($\mathcal{I} \propto t^{-\zeta \text{log}(t)}$), where $\zeta$ is the relaxation exponent. In both plots, error bars denote the error of the mean from six individual experimental realizations. All times are in units of the tunneling time, $\tau = h/(2 \pi J)$.}
	\label{tt}
\end{figure*}

\paragraph{\textbf{Experiment and Model:}} Our system is composed of a degenerate ${}^{40}$K Fermi gas prepared in an equal two-component spin mixture of its two lowest hyperfine states. The spinful fermions hop on a square lattice and the two species interact via onsite interactions that are tunable by a Feshbach resonance. Two quasi-periodic potentials with different incommensurabilities are created along the $x$- and $y$-directions of the lattice and form a quasi-periodic two-dimensional disorder potential, see \fig{schematic}. Our system is described by the following Hamiltonian:

\begin{align}
\hat{H} =&-J\sum_{\langle \textbf{i},\textbf{j}\rangle,\sigma}(\hat{c}^{\dagger}_{\textbf{j},\sigma}\hat{c}_{\textbf{i},\sigma} + \text{h.c.})\nonumber  +U\sum_{\textbf{i}} \hat{n}_{\textbf{i},\uparrow}\hat{n}_{\textbf{i},\downarrow}\nonumber\\&+\Delta\sum_{\textbf{i},\sigma}[\cos (2\pi\beta_{x} m) + \cos (2\pi\beta_{y} n)]\hat{n}_{\textbf{i},\sigma}.
\label{total_hamiltonian}
\end{align}
Here, $\hat{c}^{\dagger}_{\textbf{i},\sigma}(\hat{c}_{\textbf{i},\sigma})$ is the creation (annihilation) operator of a fermion with spin $\sigma \in \{\ket{\uparrow},\ket{\downarrow}\}$ on a lattice site $\textbf{i} = (m,n)$, characterized by the Cartesian coordinates $(m,n)$,  and $\hat{n}_{\textbf{i},\sigma} = \hat{c}^{\dagger}_{\textbf{i},\sigma}\hat{c}_{\textbf{i},\sigma}$ is the particle number operator. In the first term, the angle brackets $\langle\, ,\rangle$ restrict the sum over nearest-neighbor sites. The tunneling matrix element is set to $J \approx h\times 300\,$Hz ($h$ is Planck's constant) and $U$ denotes the on-site interspecies interaction strength. The disorder potential is characterized by the strength $\Delta$ and the incommensurable wavelength ratios $\beta_{x} \approx 0.721$ and $\beta_{y} \approx 0.693$~\cite{SOMs}. 
 
To probe the many-body dynamics of this system, we prepare a far-from-equilibrium initial state where atoms are selectively loaded only on the even stripes, see \figc{schematic}{(a)}. In an ergodic time evolution, this density-wave pattern will quickly vanish as the dynamics erase the microscopic details of the initial conditions. In contrast, a persistent pattern indicates a memory of the initial state and hence non-ergodic behavior. This can be captured by the normalized atom number difference between the even $N_e$ and odd $N_o$ stripes, defined as the imbalance $\mathcal{I} = (N_e - N_o)/(N_e + N_o)$, which serves as our dynamical order parameter.
Such an observable has several key advantages: Whereas mass transport is a slow process even in clean ergodic systems~\cite{Schneider12}, the imbalance relaxes within a few hopping times~\cite{Trotzky08,Schreiber15}. This allows us to clearly identify any longer relaxation timescales induced due to disorder. Furthermore, the \emph{dynamical} time evolution of the imbalance could capture eventual microscopic Griffiths-type effects, even in higher dimensions, where mass transport might not be sensitive to them~\cite{gopalakrishnan_griffiths_2016}. 

\paragraph{\textbf{Identifying Slow Relaxation:}} We choose a fixed intermediate interaction strength of $U=5\,J$ and monitor the time evolution of the imbalance for varying disorder strengths $\Delta$, see \fig{tt}. In the initial state, almost all the atoms occupy even stripes, such that the imbalance at zero evolution time is close to unity (see \figc{tt}{(a)}). For low disorder strength ($\Delta = 1\,J$) we observe a quick relaxation and the imbalance vanishes within a few tunneling times. However, upon increasing the disorder, relaxation slows down dramatically ($\Delta = 4\,J$) and essentially comes to a full stop for strong disorder ($\Delta = 10\,J$).

To quantitatively analyze this slow relaxation, the time dependence of imbalance is modeled as $\mathcal{I}(t) =  \widetilde{\mathcal{I}}(t) \times f(t)$. Here, $\widetilde{\mathcal{I}}(t)$ is the closed-system imbalance describing the dynamics of a perfectly isolated system, and $f(t)$ represents a weak coupling to the environment. Such couplings are present in all real systems and will always thermalize any system at long enough times~\cite{Bordia16,lueschenps16}. In our experiment, this weak coupling is dominated by a small but non-zero hopping rate between multiple two-dimensional planes along the $z$-direction, with a rate $J_\perp \approx J/10^3$~\cite{Bordia16}.
We model the resulting imbalance relaxation due to this weak coupling with a stretched exponential $f(t) = e^{-(\Gamma t)^{\beta}}$, with the decay rate $\Gamma = \Gamma_{\text{exp}} = 10^{-3}\,\tau^{-1}$ and the stretching exponent $\beta = 0.6$ measured independently in a previous experiment~\cite{Bordia16}.

\begin{figure}
	\centering
	\includegraphics[width=1.0\linewidth]{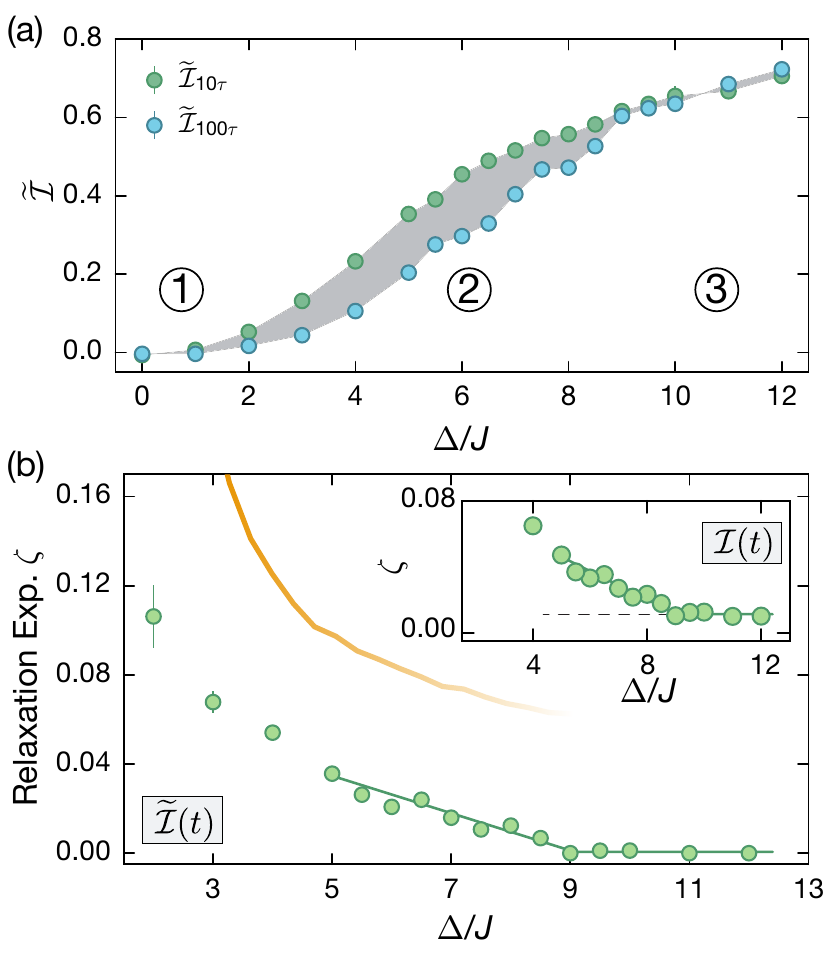}
	\caption{Closed-system imbalance and relaxation exponents near the critical regime. (a) Closed-system imbalance $\widetilde{\mathcal{I}}(t)$, at short times ($10\,\tau$, green points) and at long times ($100\,\tau$, blue points) for fixed interaction strength of $U=5\,J$. Three dynamical regimes can be identified: {\protect\circled{1}} a rapidly thermalizing regime at low disorder, {\protect\circled{2}} a regime of slow relaxation at intermediate disorder (visualized by the gray area), and {\protect\circled{3}} a regime at strong disorder in which relaxation is absent on experimentally accessible time scales, consistent with a many-body localized phase. (b) The relaxation exponent $\zeta$ obtained from \eq{eq:expo} decreases continuously with increasing disorder until a threshold value $\Delta_c$, at which it vanishes (green points). A piecewise fit $\zeta \propto \{ |\Delta-\Delta_c|^\nu, 0\}$ for $\zeta$ in the regime $\Delta \in \{5\,J , 12\,J\}$ yields $\Delta_c = 9\,J \pm 0.5\,J$ as the critical disorder strength of the possible MBL transition. The yellow line is an approximate upper bound for the relaxation exponent. The inset shows the same analysis for $\mathcal{I}(t)$, i.e., neglecting the weak coupling to the environment. Apart from a small offset (dashed line) consistent with the known background coupling~\cite{Bordia16}, we still find a sharp change of the extracted exponents at $\Delta_c$. Error bars denote the fit error, and are typically smaller than the symbol size.}
	\label{exponent}
\end{figure}
The resulting $\widetilde{\mathcal{I}}(t)$ is shown in \figc{exponent}{(a)} for short (10$\,\tau$) and long ($100\,\tau$) evolution times and fixed interaction strength $U=5\,J$ as a function of the disorder strength. We can identify three dynamical regimes. For weak disorders $\Delta \lesssim 2\,J$ \circled{1}, we observe vanishing values of short and long time imbalances, signaling the presence of a rapidly thermalizing system.
Upon increasing the disorder strength, for $2\,J\lesssim \Delta \lesssim 9\,J$, we find a regime \circled{2} characterized by a non-vanishing imbalance and clearly visible differences between the short and long time closed-system imbalances. This indicates that, in this regime, the system relaxes much slower than the microscopic timescales. For $\Delta>2\,J$, all single-particle states are localized, but in many regions of the system, interactions with nearby atoms can still result in local thermal equilibrium. However, in some rare regions with anomalously low density or large spin imbalances (see below), this thermalization mechanism could be largely ineffective. Such regions can be thermalized only by their greater surroundings, which are thermal, but to which they couple significantly weaker. Thus, the overall thermalization timescale grows.
As the disorder is increased, these surroundings themselves gradually become more localized and less effective thermal baths.
For strong disorder $\Delta \agt 9\, J$, we identify regime \circled{3}, where the values of $\widetilde{\mathcal{I}}(t)$ at short and long times are both large and, within the experimental uncertainty, identical. 
This is consistent with the system being many-body localized.

\paragraph{\textbf{Relaxation Exponents and Non-Interacting Inclusions:}} 
Identifying a suitable model to analyze the slow relaxation in regime \circled{2} is challenging, as the underlying dynamics in two-dimensions at intermediate disorders is theoretically unknown. In one dimensional models with random disorder, anomalously strongly disordered regions have been argued to give rise to a characteristic sub-diffusive phase via Griffiths effects~\cite{Agarwal15, Potter_Universal_2015, Vosk_Theory_2015, gopalakrishnan_griffiths_2016}. Because our system contains quasi-periodic disorder instead of true randomness, it should not contain such anomalously disordered regions. One possible mechanism explaining the slow relaxation at the observed timescales could be the randomness in the initial state, which can contain rare \emph{configurations}, where some regions or inclusions are effectively noninteracting and therefore, might appear localized at intermediate times, provided that the single-particle states are localized ($\Delta>2J$). In particular, thermalizing collisions cannot occur in a region where all atoms have the same spin; a region consisting, e.g., only of spin-up atoms and vacancies is effectively noninteracting and consequently, localized on intermediate timescales. These regions can relax on longer timescales by coupling to the rest of the system, provided the disorder is not too high, by two possible mechanisms: First, particles inside the inclusion can tunnel out of the inclusion, at a rate that is exponentially small in their distance from the inclusion's edge. Second, the spin and density imbalances can relax through slow nonlinear diffusion processes, so that the inclusion evaporates from outside. While this could dominate at late times, we do not expect these diffusion processes to becomes relevant on the probed timescales~\cite{SOMs}.

We define inclusions of size $L$ or larger as those containing a site from which \emph{any} path of $L$ steps encounters only atoms of one spin species or vacancies. 
The time to thermalize such a region through tunneling increases exponentially with $L$, as $t(L) \sim e^{2 L/\xi}$, where $\xi$ is the single-particle localization length of the noninteracting Aubry-Andr\'e model and the factor of two in the exponent results from squaring the matrix element in Fermi's Golden Rule. The density of such inclusions is exponentially small in their volume, i.e., $n(L) \sim p^{L^d}$ in $d$ dimensions, where $p$ is the probability of a given site belonging to such an inclusion. Combining these expressions for $n(L)$ and $t(L)$, we find that the density of inclusions that have not relaxed at time $t$ goes as $e^{-\zeta \log^d (t)}$~\cite{gopalakrishnan_griffiths_2016}, where $\zeta$ is the relaxation exponent quantifying the relaxation. While this Ansatz results in a power-law relaxation in one-dimension~\cite{agarwal_rr_16}, in two dimensions, $d=2$, it results in a slightly faster than power law relaxation of the form:

\begin{equation}
\widetilde{\mathcal{I}}(t) \sim e^{-\zeta \, \log^2(t)} \sim t^{- \zeta \, \text{log}(t)}.      
\label{eq:expo}
\end{equation}

Fitting this model to the experimental time traces in \figc{tt}{(b)} (solid lines), we find that it is able to capture the dynamical relaxation remarkably well in the entire intermediate regime \circled{2} and hence, can be used to extract the relaxation rates relevant for the probed timescales. The fitted relaxation exponents $\zeta$ are found to continuously decrease with increasing disorder strength, as shown in \figc{exponent}{(b)}, demonstrating that the system takes increasingly long to relax. Furthermore, the fitted exponent $\zeta$ appears to vanish completely beyond a critical disorder $\Delta_c$. A simple power-law fit near the critical region $\zeta \propto |\Delta - \Delta_c|^{\nu}$ for $\Delta < \Delta_c$ and 0 otherwise yields a critical disorder strength $\Delta_c = (9 \pm 0.5)\,J$ and a critical exponent $\nu \approx 0.9$. However, as the size of the critical region where universal scaling is expected to hold is unknown, the uncertainty in extracting such a critical exponent is also unknown.

Extrapolating from one-dimension, where vanishing of this relaxation exponent at a critical disorder strength $\Delta_c$ signals a transition into an MBL phase~\cite{SOMs,Vosk_Theory_2015,  Potter_Universal_2015,agarwal_rr_16,Luitz16,lueschencrit16}, our observations would support a transition into an MBL phase in \textit{two} dimensions in the isolated limit. The extracted critical disorder strength of $\Delta_c^{2\text{D}} \approx 9\,J$ is significantly larger than the corresponding critical disorder strength in one-dimension $\Delta_c^{1\text{D}} \approx 4\,J$~\cite{lueschencrit16}, even though the non-interacting critical disorder $\Delta_c^{U=0} = 2\,J$ is identical in both cases. To show that these main results are essentially independent of our model for the environmental coupling, a corresponding analysis of $\mathcal{I}(t)$, i.e. neglecting the weak coupling to the environment~\cite{SOMs,lueschencrit16}, is shown in the inset of \figc{exponent}{(b)}. The resulting relaxation exponents also continuously decrease and exhibit a sharp change before becoming constant at the same critical disorder strength $\Delta_c$, thereby highlighting the robustness of the result. The finite offset is consistent with the known coupling to the environment (dashed line)~\cite{Bordia16,lueschencrit16}.

\paragraph{\textbf{Estimating the Relaxation Exponents:}} We can employ the above-mentioned model of counting the expected non-interacting inclusions to obtain a simple theoretical estimate of the exponents $\zeta$. To this end, we assume that the surroundings of the inclusions act as a good thermal bath. To be specific, we focus on the central part of the system, in which even rows essentially reach unit filling.
Here, a site in an occupied row has probability $1/2$ of hosting say a spin-up atom; meanwhile, sites in empty rows are automatically part of the ``inclusion''. Accounting for the geometry of the inclusion, we estimate the exponent  
$\zeta_\text{th}=(\xi/2)^2\log(2)$~\cite{SOMs}. Note that this estimate does not account for imperfections of the initial state preparation~\cite{SOMs}, the inhomogeneity of the experimental system or for the possibility that an inclusion can contain some small density of particles in the other spin state without causing it to thermalize. This theoretical estimate of $\zeta$ (as indicated by the yellow line in \figc{exponent}{(b)}) could thus be regarded as an \emph{upper bound} on the relaxation exponent. It carries qualitatively similar features as the experimental data and, except close to $\Delta_c$, is also of the same order of magnitude as the experimentally extracted exponent. Near the transition, the experimentally measured relaxation exponent becomes significantly \textit{smaller} than our theoretical estimate. Such a discrepancy is expected, as in this regime the typical surroundings of an inclusion themselves become increasingly localized and thus cease to act as a good bath, which invalidates the assumptions of our theoretical model.

\begin{figure}
	\centering
	\includegraphics[width=1.0\linewidth]{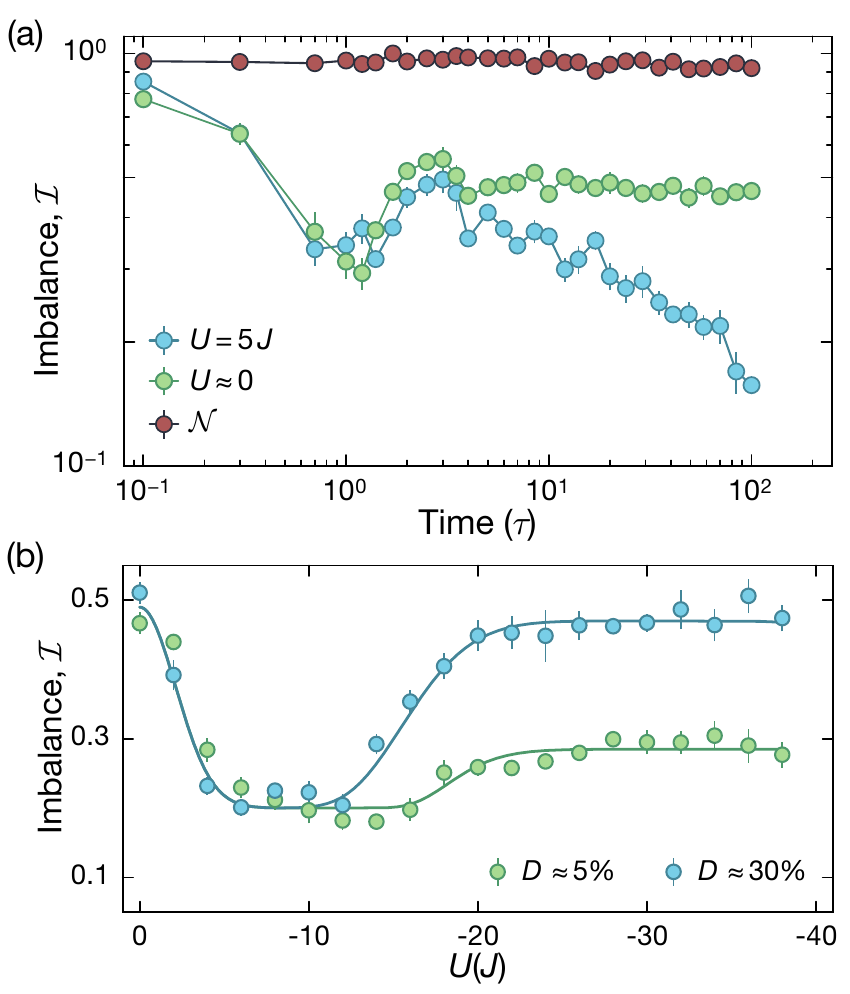}
	\caption{Interaction and energy density dependence. (a) Time evolution of the imbalance for fixed disorder strength $\Delta = 5\,J$ and vanishing $U \approx 0$ and finite interactions $U=5\,J$. While there is negligible relaxation in the non-interacting case, the interacting time trace shows slow relaxation. This  indicates that the relaxation is inherently due to interactions. We also show the normalized total atom number $\mathcal{N}$ to show that atom loss is minimal on the timescales of the experiment~\cite{Bordia16}. Each point is averaged over six individual experimental realizations and error bars denote the error of the mean. (b) We measure the imbalance for fixed disorder strength $\Delta = 6\,J$ but starting from initial states with two different doublon fractions $D \approx 5\%, 30\%$.  In the large interaction limit, we observe a significantly higher imbalance for the larger doublon fraction. This can be qualitatively understood by the reduced mobility of the bound doublon state. Solid lines are guides to the eye. Each point is averaged over six individual experimental realizations.}
	\label{pdu}
\end{figure}

\paragraph{\textbf{Interaction Effects and Energy Density Dependence:}} 
To highlight that the observed slow relaxation at intermediate disorders is driven by interactions and is completely absent in the non-interacting system, we compare it to the case of vanishing interactions $U \approx 0$ in \figc{pdu}{(a)}. While the non-interacting system is strongly localized, showing a saturation of the imbalance at a finite value, the interacting system relaxes slowly. This supports that the slow relaxation is completely interaction driven.

To probe the impact of changing the \textit{energy-density} on the relaxation, we additionally prepare initial states with finite fraction $D$ of particles on doubly occupied lattice sites (doublons), which changes the energy density of the initial state by $\sim D\times U/2$. \figc{pdu}{(b)} shows the resulting imbalance after an evolution time of $50\,\tau$. Due to a dynamical symmetry of the Hamiltonian, we expect the evolution to be the same for both repulsive and attractive interactions~\cite{Schreiber15, Schneider12}. Hence, we concentrate on the attractive side of the Feshbach resonance~\cite{Regal03} to be able to access very strong interactions. For small and moderate interactions ($|U| \lesssim 10\,J$), we measure a strong decrease in the imbalance, highlighting again the delocalizing effect of interactions. Additionally, the imbalance is identical for both energy densities. For stronger interactions, however, we find a higher imbalance for the state with more doublons, indicating a stronger localization. Qualitatively, this effect can be understood by considering the reduced hopping rate of doublons $J_D \sim J^2/U$ at large interactions, which should result in stronger localization~\cite{Schreiber15}. In one dimension, the hard core limit of infinitely strong interactions in the absence of doublons can be mapped back onto a non-interacting model~\cite{Schreiber15,jwtransform28}. Consequently, the imbalance for low doublon fraction is identical for the two extreme cases of vanishing and hard-core interactions~\cite{Schreiber15}. In contrast, such a mapping does not exist in two-dimensions and, accordingly, the imbalance is significantly different for the two extremes; again a striking qualitative difference in comparison to one dimension.

\paragraph{\textbf{Conclusion:}} We observe an extended regime of exceedingly slow relaxation of an imprinted density pattern, in which the relaxation becomes progressively slower for increasing disorder strength. After accounting for a known weak coupling to the environment, the relaxation vanishes above a critical disorder strength, thereby indicating the existence of an MBL phase in two-dimensional quasi-periodic systems. We describe a simple model based on configurational inhomogeneities in high-temperature states that captures the qualitative trend of the experiment. However, a full quantitative description of the regime of critically slow relaxation and the apparent MBL transition goes substantially beyond the currently known theory, thereby underlining the importance of experimental results in resolving such regimes. Already in one dimension, recent numerical calculations have supported that sub-diffusion can arise, at least to intermediate times, even in systems with no rare-regions in the underlying potential, such as quasi-periodic systems~\cite{barlev17,Lee17} and have shown that the emerging relaxation can appear very similar to the case of systems with a truly disordered potential~\cite{Luitz16,agarwal_rr_16}. Another possibility in two dimensions would be an intermediate critical phase between the fully ergodic and the fully MBL phase, which could correspond to this extremely broad regime with a markedly slow relaxation. 

While our results already provide important insights into MBL in higher dimensions, revealing the universal critical properties of the transition remains a challenging task for future experiments. This would require identifying the regime where critical scaling holds and to understand the role of rare-regions in quasi-periodic versus real-random systems in determining the critical properties~\cite{Khemani17,Luitz16,Lee17}. Further isolating the system would allow us to experimentally access even longer timescales and understand the coupling to an external bath~\cite{Bordia16, lueschencrit16, lueschenps16}, as well as to probe the interplay between the spin and charge sectors arising from SU(2) symmetry~\cite{ppspins16}, and plaquette resonances in two-dimensions~\cite{soonwon17}. Supplementing such results with frequency resolved spectroscopy should further provide fundamental insights into the MBL transition~\cite{BordiaFMBL16}.

\paragraph{\textbf{Acknowledgments:}}  We are grateful to David Huse and Soonwon Choi for multiple stimulating discussions. We also thank Jae-yoon Choi and Mikhail Lukin for discussions. We thank Sean Hodgman and Michael Schreiber for technical help on the 2D lattice setup. We acknowledge support from the European Commission (UQUAM, AQuS), the Nanosystems Initiative Munich (NIM), the Technical University of Munich -- Institute for Advanced Study, funded by the German Excellence Initiative and the European Union FP7 under grant agreement 291763, and the DFG grant No. KN 1254/1-1.

\bibliographystyle{nphys}
\bibliography{2D_MBL_Arxiv_v1}
\cleardoublepage

\appendix

\setcounter{figure}{0}
\setcounter{equation}{0}

\renewcommand{\thepage}{S\arabic{page}} 
\renewcommand{\thesection}{S\arabic{section}} 
\renewcommand{\thetable}{S\arabic{table}}  
\renewcommand{\thefigure}{S\arabic{figure}} 
\renewcommand{\theequation}{S\arabic{equation}} 

\section{\Large{SUPPORTING MATERIAL}}
\setlength{\intextsep}{0.8cm} 
\setlength{\textfloatsep}{0.8cm} 

\section{\large{Data Analysis and Supporting Measurements}}
\begin{figure}[b!]
	\centering
	\includegraphics[width=1.0\linewidth]{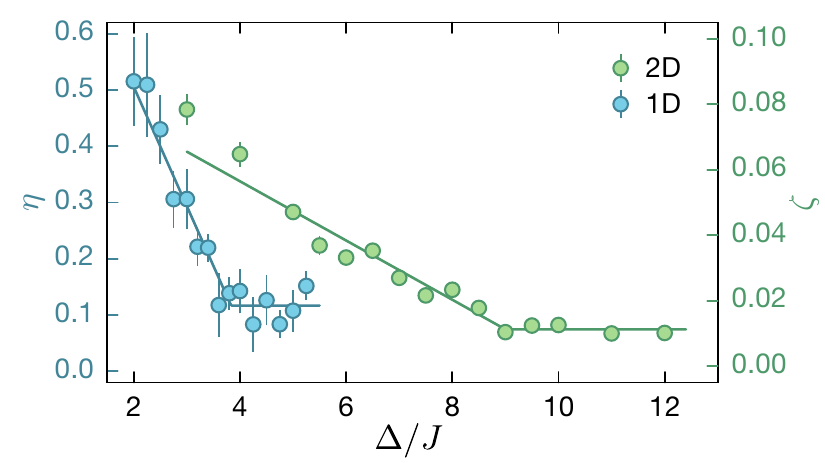}
	\caption{Relaxation Exponents in 1D vs. 2D: We compare the power-law exponent $\eta$ obtained in 1D from fits $\mathcal{I}(t) \propto t^{-\eta}$~\cite{lueschencrit16} to the relaxation exponent $\zeta$ obtained in 2D (\fig{exponent}{(b)}). The graph shows the strong effect of dimensionality, with a much higher critical disorder strength in two-dimensions $\Delta_c^{\text{2D}} \approx 9\,J$ as compared to the one-dimensional case $\Delta_c^{\text{1D}} \approx 4\,J$, which demonstrates the additional ways for delocalization in higher-dimensions. Solid lines are guides to the eye and error bars denote the fit error. The interaction strength is $U=4\,J$ in 1D and $U=5\,J$ in 2D case. Hence, the resulting MBL transition in 2D gets as much support as in 1D, yet it is located at a markedly larger value of the critical disorder strength.}
	\label{plexps}
\end{figure}

\paragraph{\textbf{1D-2D Comparison:}} We compare the power-law relaxation exponent $\eta$ from of the 1D system (reported in Ref.~\cite{lueschencrit16}) to the relaxation exponents $\zeta$ obtained in this work in 2D in \fig{plexps}. The abrupt change of the behavior of the relaxation exponents at a certain value of the disorder strength indicates the putative MBL transition. We note that the different magnitude of the exponents $\eta$ in 1D and $\zeta$ in 2D results from extracting using different functional forms. Hence, the graphs shows that for almost all the disorder strengths where slow relaxation is observed in 2D, the 1D system is already in the MBL phase. This highlights the consequences of dimensionality on the relaxation dynamics and shows the very different critical disorder strengths.

\begin{figure}[b!]
	\centering
	\includegraphics[width=1.0\linewidth]{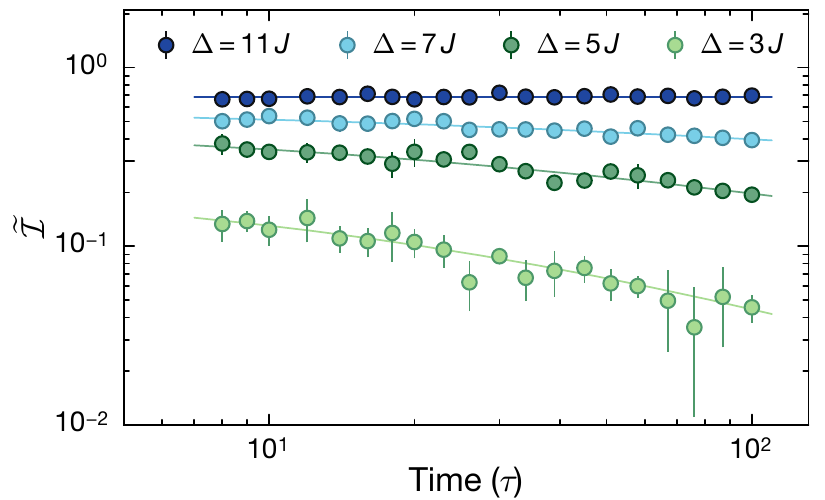}
	\caption{Closed-system imbalance as a function of time: We show \cohimb{} as a function of evolution time to see the slow relaxation, exhibiting a downward bending on the double logarithmic scale. The data is for the same parameters as in \fig{tt}, but corrected for the finite background coupling $\Gamma = \Gamma_{\text{exp}}$. Solid lines are fits as described in the main text.}
	\label{ttcohimb}
\end{figure}

\paragraph{\textbf{Power Law Fits:}}  We plot the closed-system imbalance $\widetilde{\mathcal{I}}$ on a double logarithmic plot (\fig{ttcohimb}). The closed-system imbalance appears to decay slightly faster than a power-law, supporting the analysis in the main text. However, to test the robustness of the main results presented in our paper, we also fit power laws $\propto t^{-\eta}$ without the logarithmic correction to the relaxation of the imbalance in 2D and find that the critical disorder strength $\Delta_c$ and the critical exponent $\nu$ remain unchanged (not shown).

\begin{figure} [t!]
	\centering
	\includegraphics[width=1.0\linewidth]{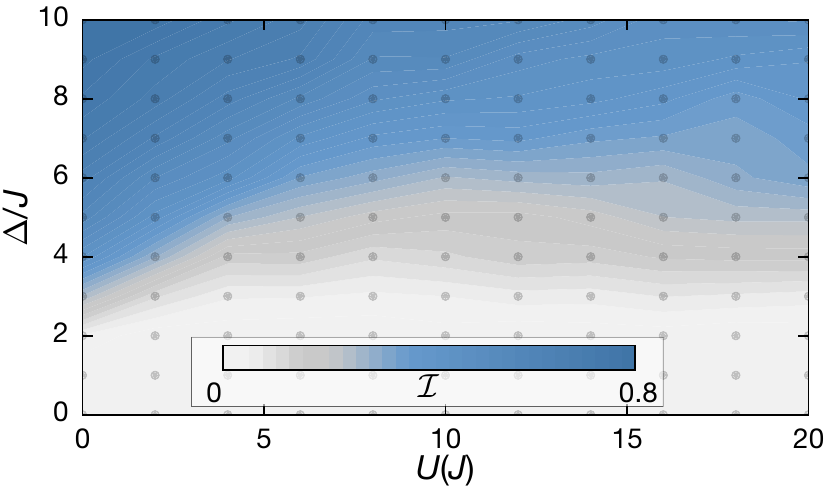}
	\caption{Interaction dependence of the intermediate time imbalance: The imbalance measured at times $\sim 50\,\tau$ shows a pronounced change as a function of both disorder and interaction strength. Generally, we find that the interactions tend to decrease the imbalance whereas disorder tends to increase it. In contrast to one-dimension, where infinitely strong interactions map to an effectively free theory, in our two-dimensional sample the large interaction limit is significantly different from the noninteracting case. Experimental data is measured at the gray dots. Each point is an average of six individual experimental realizations.}
	\label{dupd}
\end{figure}

\paragraph{\textbf{Interaction Dependence of Intermediate Time Imbalance:}}
Our experiment also allows us to explore the influence of the \textit{interaction strength} on the relaxation dynamics by using a Feshbach resonance~\cite{Regal03}. We measure the imbalance at finite but late evolution times of $50\,\tau$ as a function of the interaction strength \fig{dupd}. We observe that tuning from weak ($U \sim 0$) to intermediate interactions ($U \sim 5\,J$) causes a significant reduction of the imbalance, highlighting the delocalizing effect of interactions. This trend stops when tuning to even stronger interactions $U > 10\,J$ where a slight increase of the imbalance is measured. 
We note that in the non-interacting limit our system decouples the two orthogonal directions and effectively reduces to a one-dimensional problem~\cite{Bordia16}. This happens because the onsite potential is a sum of two quasi-periodic terms which can be prepared along two orthogonal directions. The solution to the two-dimensional system in the non-interacting case is therefore the same as the solution of the one-dimensional problem and the transition stays at $\Delta_c = 2\,J$, as explored in our earlier works ~\cite{Schreiber15, lueschencrit16}. We also check this explicitly by comparing the imbalance after an evolution time of $50\tau$ in the non-interacting case in one and two-dimensions and find them to be identical.

\paragraph{\textbf{Estimating Effects of a Weak Coupling to the Environment:}} We understand the effects of a weakly coupled environment in our system from our recent works in one-dimension~\cite{lueschencrit16} and from coupling one-dimensional tubes~\cite{Bordia16}. In the latter, the hopping between multiple one-dimensional tubes enables them to collectively act as a bath for each other. Numerical simulations in one-dimension provide strong support that at times which are at least about a few hundred tunneling times, the effect of the bath can be well captured by a multiplicative stretched exponential as discussed in the main text~\cite{lueschencrit16}. To study the robustness of our conclusions, we perform fits using the stretching exponents $\beta$ between $0.4-0.7$ and the background lifetime $1/\Gamma$ between $900-1200 \,\tau$ which leaves the main results unchanged (not shown). We note that even completely neglecting the weak-coupling to the bath gives a sharp change of the relaxation exponent around the same critical disorder strength $\Delta_c$ and has no effect on the observation of slow dynamics, as shown in the time traces themselves.

\begin{figure}[tb!]
	\centering
	\includegraphics[width=1.0\linewidth]{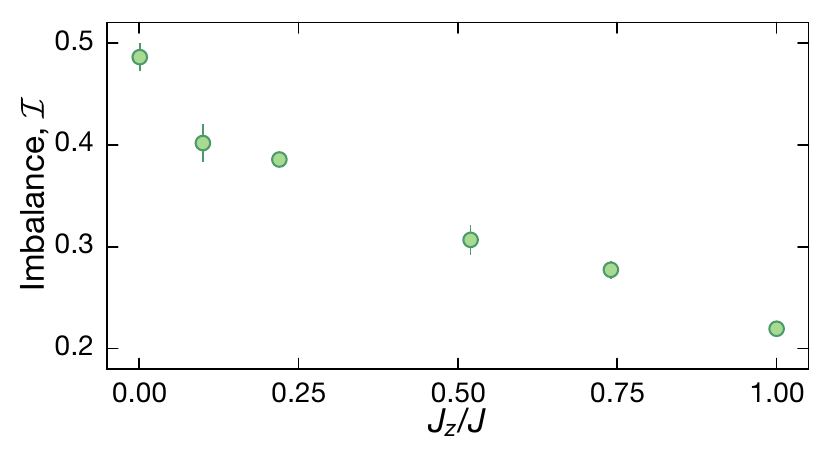}
	\caption{Coupling identical planes: The graphs shows the measured imbalance as a function of hopping  $J_z$ in the $z$-direction. It is measured after an evolution time of $50\,\tau$ for fixed disorder strength $\Delta = 8\,J$ and interactions $U = 5\,J$. Similar to our previous work on coupling identical many-body localized systems in one-dimension~\cite{Bordia16}, we indeed find that for higher coupling strength between the two-dimensional planes (and hence larger $J_z/J$) the imbalance reduces. Note that the smallest value of $J_z/J$ in the figure is $10^{-3}$, as used in the main text, and not exactly zero. Each point is an average of six individual experimental realizations and error bars denote the error of the mean.}
	\label{couplingplanes}
\end{figure}

\paragraph{\textbf{Coupling Identical Two-Dimensional Planes:}} We explicitly check that indeed the coupling between different two-dimensional planes can result in faster delocalization, as in our previous work on coupling one-dimensional MBL systems~\cite{Bordia16}. We measure the imbalance after $50\, \tau$ as a function of hopping in the $z-$direction, see \fig{couplingplanes}, and find that indeed larger couplings (i.e. larger $J_z/J$) results in lower imbalances. All measurements are performed for fixed interactions $U = 5\,J$ and disorder strength $\Delta = 8\,J$.

\section{\large{Theoretical Analysis}}

\begin{figure*}
	\centering
	\includegraphics[width=1.0\linewidth]{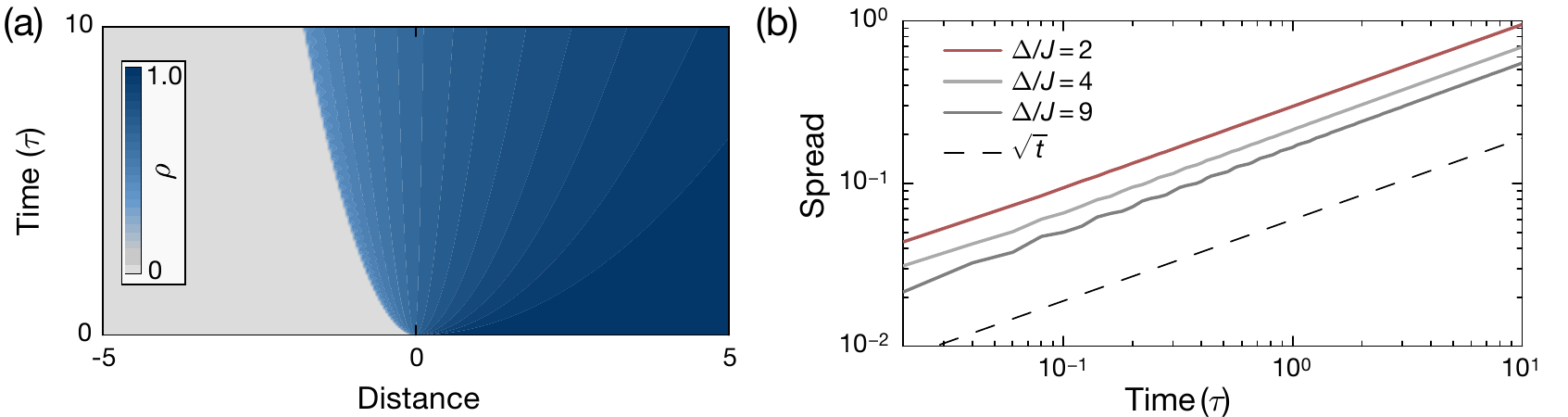}
	\caption{Numerical solution of the nonlinear diffusion equation. (a) The nonlinear diffusion equation \eqw{nldiff} is solved numerically for an initial step-density profile ($\rho(x)=0$ for $x<0$ and $\rho(x)=1$ for $x>0$) and $\Delta/J=4$. (b) The spread of the wavefront into the initially empty region ($x<0$) is diffusive and hence grows as $\sqrt{2\alpha t}$. With increasing disorder strength the effective diffusion constant $\alpha$ decreases. }
	\label{fig:spread}
\end{figure*}

\emph{\textbf{Single-Particle Localization Length $\xi$:}} Our theoretical estimate of the relaxation exponent $\zeta_\text{th}=(\xi/2)^2 \log(2)$ requires the localization length $\xi$ of the noninteracting system. Assuming exponentially localized single-particle states $\psi(\mathbf{r})=e^{-|\mathbf{r}|/\xi}$, we obtain the non-interacting localization length $\xi$ by relating it to the inverse participation ratio $\text{IPR}=\frac{1}{N^2} \sum_\mathbf{r} |\psi(\mathbf{r})|^4$, where $N$ is the wavefunction normalization $N=\sum_\mathbf{r} |\psi(\mathbf{r})|^2$. The inverse participation ratio can be computed numerically from exact diagonalization of the non-interacting system, as described by \eq{total_hamiltonian} with $U=0$. We  average the inverse participation ratio over different phases of the quasi-periodic potential and choose systems of size $48 \times 48$ to obtain results that are representative for the thermodynamic limit. 

\emph{\textbf{Size and Counting of Inclusions:}} The size of an inclusion is operationally defined as follows (for concreteness we assume an inclusion consisting of spin-up and empty sites): if, starting from a particular site $\mathbf{x}$, all paths of $L$ nearest-neighbor steps encounter only spin-up or empty sites, then $\mathbf{x}$ belongs to an inclusion of size $\geq L$. Note that the total matrix element from $\mathbf{x}$ to a point $\mathbf{y}$ at the edge of the inclusion includes a combinatorial factor from the number of paths connecting the two points. However, provided that single particle states are well localized, this combinatorial factor is dominated by the exponential decrease in the matrix element as a function of $L$.

Thus, the minimal inclusion of $ L$ sites centered at $\mathbf{x}$ is a square, with diagonals of length $2L$ along the lattice directions. The area of the inclusion is then $2L^2$. If the initial state created in the experiment were purely random, the probability of such an inclusion would be $p^{2L^2}$, where $p$ is the probability of a spin-up or empty site. However, since the initial state in the experiment ideally consists of alternating rows of empty and filled sites, the true probability of an inclusion is $p_f^{L^2}$, where $p_f=1/2$ is the probability that a site in one of the \emph{occupied} rows is either spin-up or empty. Setting $L = \xi/2 \log t$ this leads to the simple parameter-free estimate in the main text. (Recall that the decay at distance $L$ is largely incoherent, and hence has a timescale set by $t \sim \exp(2L/\xi)$, where the factor of two results from Fermi's Golden Rule.)

We note that this simple estimate systematically \emph{undercounts} inclusions. More refined estimates would allow for the presence of a sufficiently small density of spin-down particles inside the inclusion, as these will not themselves suffice to thermalize the inclusion. Moreover, our counting of inclusions considered fully occupied even rows and hence focused on the central region of the trap, where the density is higher and inclusions are correspondingly rarer. 

\emph{\textbf{Nonlinear Diffusion and the late-time ``filling-in'' of Inclusions:}} The ``tunneling'' process out of the inclusions is one of two possible mechanisms by which an inclusion can relax. The other mechanism is for particles to diffuse in from the edge of the inclusion. As we now discuss, this diffusion process dominates tunneling at sufficiently long times. 

We note that for $\Delta > 2J$, single-particle states are localized. Thus, in the low-density or highly spin-polarized limit, the diffusion constant must vanish. Indeed, the diffusion constant should be either strictly zero or nonperturbatively small below some critical (or crossover) density $\rho_c$. We can estimate the density-dependence of the diffusion constant perturbatively as follows. Since single-particle moves do not cause relaxation, relaxation must be due to collective $m$-particle rearrangements. The matrix element for such rearrangements decreases as $\sim J \exp(-m/\lambda)$, where $\lambda$ is a parameter that decreases with increasing disorder; the level spacing for such rearrangements also decreases as $\Delta \exp(-s m)$, where $s \sim \rho \log 4$ is an entropy per site, and $\rho$ is the density per site. When $\lambda s > 1$ the system is in the thermal phase, and the dominant rearrangements are at the scale $m^* \sim \log(\Delta/J)/(\rho \log 2 - 1/\lambda)$, i.e., they involve rearranging $m^*$ particles. The diffusion constant then depends on the density as $D(\rho) \sim \rho^{m^*}$. 

The general nonlinear diffusion equation in this system is of the form 

\begin{equation}\label{nldiff}
\frac{\partial \rho}{\partial t} = c \nabla \left( \rho^{m^*(\rho)} \nabla \rho \right).
\end{equation}
where $c$ is a density-independent prefactor. 
Starting with a step profile, the diffusion into an empty region from a region of typical density can be seen numerically to be diffusive (\fig{fig:spread}); however, the diffusion constant of this ``front'' is strongly suppressed relative to that inside the typical region. Assuming diffusive behavior, the timescale for an inclusion of size $L$ to thermalize is $t \sim L^2$; adapting the counting argument above, we expect that in this case the decay due to rare regions will be exponential in time. Hence, the ``filling-in'' of inclusions dominates the ``tunneling'' process out of the inclusions at late times.

\emph{\textbf{Crossover from Griffiths to Diffusive Behavior:}} At present, it is unclear whether the perturbative estimates above are valid near the MBL transition. Nevertheless, we can use these estimates to identify a timescale on which Griffiths effects should cease to dominate dynamics in incommensurate systems. This timescale is estimated as follows. The rate at which a particle flips by ``tunneling'' out of an inclusion is exponential in the size of the inclusion; however, the phase space for a tunneling event grows as the boundary area of the inclusion. Thus, for an inclusion of size $L$, the tunneling rate is $\sim J\exp(-2L/\xi) L$ as obtained from Fermi's Golden rule. Meanwhile, the dominant \emph{typical} relaxation outside the inclusion occurs at some, \emph{a priori} unknown rate $\gamma$, governed by some optimal $m$-particle rearrangement. This leads to an effective diffusion time of order $L^2 /(\alpha \gamma)$ for an inclusion of size $L$, where $\alpha \approx 10^{-2}$ is the effective diffusion constant obtained from numerically solving the nonlinear diffusion equation \eqw{nldiff}. The crossover occurs for inclusions of size $J\exp(-2L/\xi) L^3 = \alpha \gamma$. Solving this equation perturbatively for $L$, we obtain $L=(\xi/2) \log[J/(\alpha\gamma)]$. From inverting the effective diffusion time $L^2/(\alpha \gamma)$, we estimate the rate at which such inclusions relax $\sim 4 \alpha \gamma/(\xi \log J/(\alpha\gamma))^2$. 
Unfortunately, we cannot directly measure the rate $\gamma$ at which typical regions relax; however, it is clear that $\gamma \alt J$. Therefore, we can bound Griffiths effects to be visible for times that are \emph{at least} on the order of $100/J$. As $\Delta$ increases, this crossover scale moves to later and later times.

\emph{\textbf{An alternative mechanism:}} We have highlighted the interpretation of the observed slow relaxation in terms of Griffiths effects, as these reproduce both the functional form and semi-quantitatively the observed prefactor. It is possible that other mechanisms might also play a role in the slow relaxation, however. One example of such an alternative mechanism is as follows: the spin exchange coupling constant, $J_s \sim U(J/\Delta)^2$, is much slower than the rate of charge motion. However, since the spin-exchange Hamiltonian has random couplings up to a weak breaking of the exact SU(2) symmetry~\cite{BordiaFMBL16}, but not random on-site fields, presumably spin excitations are substantially less localized than charge. Thus, one possible model of the intermediate regime is in terms of localized charge excitations coupled to a spin bath~\cite{pg2017}. Spin exchange processes alone do not relax the charge imbalance; that requires charge to move, using the spin as a bath with a narrow bandwidth or a long correlation time~\cite{gn, pg2017}. In this situation, the spin-mediated charge rearrangement rate is parametrically smaller than the spin timescale, being given by $\Gamma_c \sim J (J_s/J)^{2/(\lambda_c \log 4)}$~\cite{pg2017}, where $\lambda_c \alt 1/\log 4$ is the localization parameter of the charge sector. This form arises because finding a charge transition that is resonant to within energy $J_s$ requires going to high orders in perturbation theory~\cite{gn}. Since $\Gamma_c \ll J_s$ this picture naturally accounts for the slowness of relaxation.

This explanation is also consistent with the existence of multiple timescales, as follows: although the \emph{typical} charge relaxation timescale is very slow, some charge modes will relax atypically fast. Specifically, let us consider the imbalance at a timescale $t$. On this timescale, $m$-particle rearrangements can occur when $J_s \exp(-2m/\lambda_c) \agt 1/t$, so $m \alt (\lambda_c/2) \log(J_s t)$. The fraction of possible rearrangements at this scale is $(J_s/J) \exp(m \log 4)$, which is (by definition) less than unity on timescales shorter than $1/\Gamma_c$. By this argument, the fraction of the system that will relax on timescale $t$ is $(J_s/J) \exp[(\lambda_c/2) \log 4 \log(J_s t)] \sim t^{\log 4 \,\lambda_c /2}$. The imbalance should decay, according to this mechanism, as $t^{-\log 4 \, \lambda_c/2}$. 
While consistent with slow relaxation, therefore, this scenario does not account for the observed dimension-dependence of the imbalance decay. 

\section{\large{Technical Details of the Experiment}}

\paragraph{\textbf{Lattice Parameters:}} All lattice potentials result from retro-reflected laser beams of wavelengths of either 532\,nm, 738\,nm or 1064\,nm. As in our previous works, (Refs.~\cite{Schreiber15,Bordia16}), overlapped lattices of 532\,nm and 1064\,nm light form a superlattice along the $x$-direction that is used to prepare the initial density wave and to read out the even-odd imbalance. An additional, weak and incommensurate lattice at 738\,nm forms the disorder lattice along the $x$-direction and the same laser is used to form the primary lattices along the orthogonal $y$ and $z$-directions. Compared to Ref.~\cite{Schreiber15}, the lattice along the $y$-direction is less deep to allow tunneling also in this direction and we have added a second, weak and incommensurate lattice along this direction using the existing 1064$\,$nm laser. This results in the two slightly different incommensurable wavelength ratios $\beta_x = 532/738.2 = 0.721$ and $\beta_y = 738.2/1064 = 0.693$.

\begin{figure}[b!]
	\centering
	\includegraphics[width=1.0\linewidth]{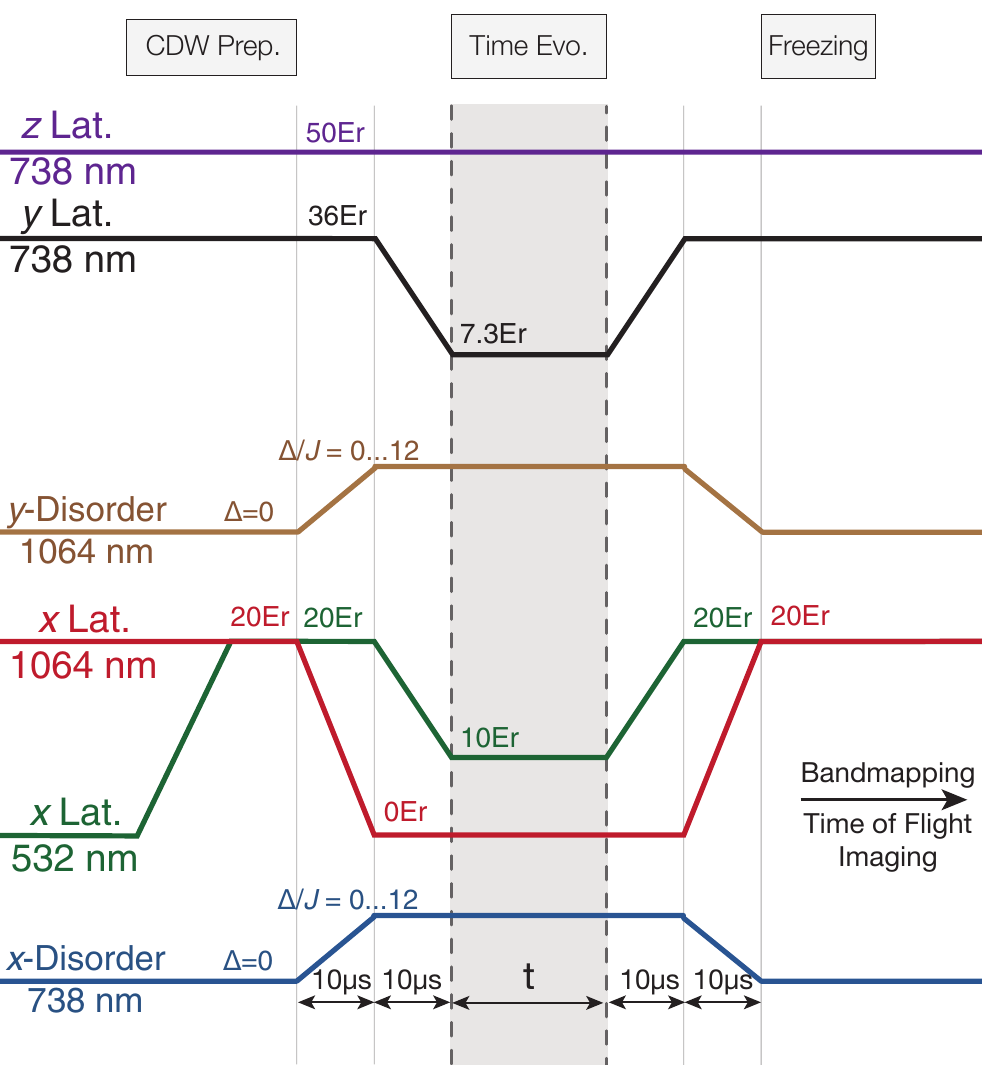}
	\caption{A schematic showing the lattice sequence for preparation and detection. Please see text for full details.}
	\label{prepdec}
\end{figure}

\paragraph{\textbf{General Sequence:}}The experiment produces an ultracold gas of fermionic Potassium-40 ($^{40}$K) atoms by sympathetically cooling $^{40}$K with bosonic Rubidium-87 ($^{87}$Rb) in a plugged quadrupole trap followed by an optical dipole trap. Reducing the dipole trap depth lower than a threshold value completely removes $^{87}$Rb due to its higher mass, such that only $^{40}$K remains in the trap. We further evaporate $^{40}$K in an equal mixture of the two lowest hyperfine states of the $^4S_{1/2}$  manifold ($\ket{F,m_F} = \ket{9/2,-9/2} \equiv \ket{\downarrow}$ and $\ket{9/2,-7/2} \equiv \ket{\uparrow}$) to a final temperature of $T/T_F = 0.19(0.02)$, where $T_F$ is the Fermi temperature. Interactions between the two states can be tuned via an s-wave Feshbach resonance centered at 202.1G~\cite{Regal03}. Except in \figc{pdu}{(b)}, the scattering length $a$ is set to $140 \, \text{a}_{\text{0}}$ during lattice loading to restrict the doubly occupied sites to a maximum of about 5\,$\%$. Here, $\text{a}_{\text{0}}$ denotes the Bohr radius.

After loading into the deep lattices, we then set the scattering length to control the desired interactions $U$ in the lattice for the following evolution time. At the end of this preparation stage, we ramp down the $x$-main lattice from $20 \, E_{\text{r,i}}$ to $10 \, E_{\text{r,i}}$ in $10 \, \mu$s and ramp down the $x$-long lattice to $0\,E_{\text{r,i}}$ in the same time. The primary lattice along the $y$-direction is simultaneously ramped down to $7.3 \, E_{\text{r,i}}$ in order to have the same hopping element $J$ in both directions. Here, $E_{\text{r,i}}$ is the recoil energy of the respective lattice $i \in {x,y}$. At the same time, the disorder lattices are also ramped up along both the $x$ and $y$ directions to the desired value so that we obtain the same value of the effective disorder strength in both directions. All of these lattice ramp times are short compared to a tunneling time $\tau$. Simultaneously, we also ramp down the net dipole confinement to have overall less than $5$\,Hz trap frequency. The system is then allowed to evolve for various evolution times. For detection, the short lattice, the long lattice and the orthogonal lattices are ramped high again to inhibit hopping and freeze all occupations. Finally, we employ a bandmapping technique to obtain the number of atoms on the even and odd sites~\cite{Foelling07}. All bandmapped images are taken after $8 \,$ms time of flight and are imaged along the $y$-axis. Further details of the bandmapping procedure, imbalance extraction and others are provided in our earlier works~\cite{Schreiber15,Bordia16}. A simplified diagram of the lattice laser ramps is shown in \fig{prepdec} for easy visualization.

\paragraph{\textbf{Disorder Strength:}}The disorder strength $\Delta$ depends on the lattice depth of the main lattice, the lattice depth of the disorder lattice and the ratio of their wavelengths $\beta_{\text{i}}$. In $x$-direction we use $10\,E_{\text{r},x}$ as the main lattice depth, and in $y$-direction we use $7.3\,E_{\text{r},y}$. Here $E_{\text{r,i}}$ is the recoil energy of the main lattice in the respective direction. The disorder strength is then given by $\Delta/J = \alpha \,s_d$, where $s_d$ is the depth of the disorder lattice in units of its recoil energy. The value of $\alpha$ in the $x$-direction is 11.11 and in the $y$-direction 5.21~\cite{Schreiber15}.

\end{document}